# *K*-Geodetic Graphs and their Applications to Analysis across Different Scales of Dynamics of Complex Systems

Carlos E. Frasser, George N. Vostrov

*Department of Applied Mathematics, Odessa National Polytechnic University, Ukraine*

This paper describes a new approach to the problem of the structural research of clusters based on the theory of geodetic and *k*-geodetic graphs. We firmly believe that this same approach can be used when solving problems of correlation between structural and spectral metrics in complex networks. So, what we want to point out is about the possibility of applying one theory of networks (the theory of *k*-geodetic networks) to the solution of problems in another type of networks (complex networks). The theory of geodetic graphs and their various modifications represents an important tool for the structural analysis of complex systems of transmission, processing, and analysis of information. In the case of large data sets, their stochastic dependence is described by large-dimensional correlation matrices. One of the problems of correlation analysis is the study of the structure of the correlation matrix. It is proved that such a structure is adequately described by geodetic graphs. The obtained structural data allow solving the choice problem of significant variables in multidimensional regression models.

## 1. The Problem of Structural Analysis of Complex Systems of Transmission, Processing, and Analysis of Information and its Relation to *K*-geodetic Graphs.

Information systems include computer systems whose software is oriented to the processing and analysis of large data sets. One of the important areas of data analysis is its correlation analysis. Information on the relationship of random variables is represented in the form of correlation matrices that are a source of information for its own correlation analysis, regression analysis, and canonical correlation analysis. In the last two cases, there are problems when choosing the significant variables of the regression models and searching for the shortest pathways between variables that explain the stochastic relationship between two given random ones. The choice of such pathways is reduced to the selection of special classes of geodetic and bigeodetic graphs. One of the problems of correlation analysis is the study of the structure of the correlation matrix. The structure of correlation matrices can be well described by different classes of geodetic graphs. So geodetic graphs and their various modifications becomes an important tool for the structural analysis of complex systems of transmission, processing, and analysis of information.

The goal of publications [4,5,7] has been to create the mathematical tools for designing information (computer) systems of optimum topological structure [6] and studying stochastic dependence structures in multidimensional correlation analysis, structures of canonical regression models, similarity matrices in automatic classification (cluster analysis) and pattern recognition. It is shown that the theory of *k*-geodetic graphs is the most appropriate tool for the solution of problems of structural analysis. The main focus is on those properties of graphs that most fully reflect the applied problems of creating information (computer) systems and developing mathematical methods of information processing and analysis. Although the literature on the methods of designing and creating information (computer) systems [2,9] as well as methods of information processing and analysis [1,3] is very common, nevertheless in these areas there are many unsolved problems that are related to the need of using the structural properties of computer systems, correlation and autocorrelation matrices, and canonical regression models. Attempts have been made to study the structures of correlation matrices in correlation analysis, but they are more descriptive in nature than intended to obtain quantitative estimates of the relationship between random parameters and the influence degree of cyclical structures on the extent of describing certain random variables throughout others. Analysis of these publications shows that the problem of structural analysis of data is not deeply developed. Therefore, it is necessary to have complete information on geodetic graphs [4], their structure [7], and also properties and methods for solving extreme problems over geodetic graphs [6].

## 2. Cluster Analysis as a Method of Information Processing Based on the Application of Geodetic Graphs.

In the theory of automatic classification, an important problem is the description of class structures. The concept of a class in automatic classification is based essentially on the notion of a cluster. At the same time, there are several meaningful interpretations of the concept of cluster. Therefore, both concepts (class and cluster) need to be clarified in their significant meaning, as well as their formalization. We will herewith proceed from the following assumptions:

   (a) in automatic classification and cluster analysis, it is assumed that the description space of the classification objects is complete,
   (b) in the description space of objects, an adequate metric or measure of similarity is chosen,
   (c) in the description space with given metric, the hypothesis of local compactness is valid,
   (d) classes in automatic classification are the union of clusters belonging to some of their decompositions,
   (e) the set of all clusters admits a partition into non-intersecting subsets that satisfy the condition of global compactness.

Information about the class in automatic classification that forms its clusters and their structures is determined by the choice of characteristics (parameters) of the object description, a given metric (or measure of similarity), the quality functional of automatic classification, and another number of other a priori assumptions.

Let us consider the meaning of the formulated assumptions and their effect on the obtained results of automatic classification. In the evaluation of the results of automatic classification, the structure of classes in the problem of information processing plays an important role and depends essentially on the set of characteristics that describe the classified objects. It is assumed that a priori a selected set of characteristics includes all the necessary information that determines the expected classification. The description is non-redundant, but the set of characteristics forming the description space includes all the indications necessary for constructing the classification expected on the basis of a meaningful formulation of the problem. Redundancy is implied in the sense that a set of characteristics contains at least one subset, the removal from the description of which generates a classification in the meaningful and formal senses. The resulting classifications are to be identical. If the description space is incomplete, then there exists a nonempty subset of characteristics whose deletion from the description worsens the meaningful interpretability of the resulting partition of objects into classes. It is obvious that the redundancy of the description space is not a sufficient condition for its completeness. Full space can be redundant. Redundancy affects the computational complexity of solving automatic classification problems, but does not change the division of objects into classes and the structure of most classes. This paper considers the problem of describing the structure of clusters and classes, and therefore, the problem of minimizing the description space by the number of attributes included in them is not considered.

The meaningful interpretability of classes and clusters and their structural properties essentially depends on the choice of the metric or the similarity measure of the classified objects. The choice of a metric or similarity measure will be called consistent with the problem of automatic classification if for a given metric, the resulting classification completely agrees with valuable considerations, a priori information, and sometimes with intuitive notions. It can be argued that the choice of the metric is more likely to affect the partitioning of the original set into classes than their structure. At the same time, the class structure is an important source of information obtained as a result of automatic classification.

Structural information plays a particularly important role in the development of optimal management decisions when creating an optimal control strategy in automated systems. The construction of an optimal control strategy is greatly simplified if in the classes with similar conditions of the controlled object, the choice of the condition in which is necessary to fulfill a transition is carried out by using information about the space arrangement of classes relative to each other, information on the internal structure of classes and the clusters that form them.

The structure of classes and clusters depends to a large extent on whether there exists such a threshold "$\delta$" for the distance between two objects in the description space, that if for any two objects the distance between which is less than $\delta$, then such objects necessarily belong to each cluster, and therefore, to the class. It is obvious from the practical importance of considerations that such a threshold must exist. If this is not true, then either the description space includes only non-significant variables, which contradicts the first assumption, or it must be agreed that objects having practical identities have completely different properties which evidence themselves in belonging to different clusters and classes. If this is true for any arbitrarily small $\delta$, then the distribution of the belonging of objects to different clusters and classes in the description space will have a disordered vague character. At the same time, from meaningful considerations, it is clear that if we take an arbitrary object and slightly change its characteristics in the description space, then the similarity of the transformed object to the original one will be close to unity. Therefore, we can assume that for the entire set of classified objects of a certain nature the principle of local compactness is valid.

**The principle of local compactness.** If a set of objects of a certain nature is given in a complete description space with a uniform metric, then there exists a max $\delta$ such that if the distance of the object from one of the objects of the selected set of objects is less than $\delta$, then such an object also belongs to the same cluster as the selected object. The principle of local compactness requires a meaningful refinement in some respects:

i) what classification properties should be possessed by the objects lying on the boundary between two clusters and classes;

ii) how the principle of local compactness is fulfilled on the boundary of clusters and classes;

iii) in what terms should the structure of classes and clusters be described so that it reflects, to a maximum extent, information about their construction and properties.

For clusters belonging to the same class, their mutual intersection or tangency is allowed. In a meaningful sense, this does not contradict the principle of local compactness since clusters belong to the same class. When considering classes, each of which is the union of some set of clusters, the principle of local compactness acts only within a certain minimally piecewise-smooth shell spanned by each class. We assume that the minimal shell of classes does not intersect. The points placed between the shells do not have a clear-cut membership in one of the classes. In the border areas between classes, the belonging to classes is vague. For its description, one can use probability models or classification methods based on the theory of fuzzy sets. The set of objects whose belonging to classes are uniquely determined constitutes the deterministic part of the classification model.

It is important to have methods for describing the structures of clusters and classes. The mathematical apparatus underlying such methods should allow a clear description of the internal structure of clusters and classes, and also, the external structure reflecting the arrangement of classes and clusters relative to each other in space. As a consequence of structural analysis, it is desirable to obtain information on the structure of the transition regions between classes.

In multidimensional spaces a complete description of the mutual allocation of points of a given set relative to each other represents an important applied problem. Its complexity lies in the choice of adequate characteristics of sets that are sufficiently complete for describing the structure. From the point of view of naive cluster analysis, the greatest interest is produced by such subsets of objects (points) that form condensations with centers of condensation. For each condensation, one can find a threshold $\delta_0$ such that the distance between each pair of objects (or points) does not exceed $\delta$. This fact is structurally described by a complete graph. From a meaningful point of view, condensations can be interpreted as clusters and the corresponding complete graphs can be viewed as a structural model of clusters. It should be noted that the requirement of the graph being complete is very restrictive; especially if the value of the compactness level δ is unknown. It is obvious that with such an approach, a pair of vertices of the graph is connected by an edge if the distance between objects is not more than a given value. As the center of condensation, you can select an object for which the average distance to the others or the variance of objects relative to the specified object is minimized. With this approach, the condensation can have several centers of condensation. The choice of centers of condensation, from a meaningful viewpoint, is not fundamental, but formally such a choice can influence the structure of the cluster, and therefore, the graph describing the cluster as well as the structure of the entire cluster diversity describing the classification structure of the whole set of objects. This is due to the fact that a lot of clusters can be considered as a set of data generators when solving problems of automatic classification.

The problem of automatic classification is reduced to the problem of dividing a set of clusters into disjoint subsets, provided that clusters belonging to different classes do not intersect. A variant of automatic classification based on the idea of class intersection is possible. This option is usually associated with stochastic classification models. The development of this approach assumes a fuzzy or stochastic model of clusters. In this paper, we confine ourselves to the deterministic case. So we assume that classes do not intersect and are separated by regions. The points of these regions correspond to objects whose belonging to classes does not have a clearly expressed character. From a meaningful point of view, the existence of separated regions is justified by the fact that the transition from one class to another has vague character.

The solution of the problem of structural analysis of clusters essentially depends on the meaningful sense of the concept of cluster and its formalization. The most meaningful and easily formulated concept is constructed on the basis of the concept of maximal complete graphs.

Let $B = \{B_1, B_2, \ldots, B_k\}$ be a set of classified objects, each of which is described in some *n*-dimensional space, which is not necessarily Euclidean. For instance, one in which each object is described in the space of attributes related either to quantitative or qualitative scales. Let us assume that a matrix or measure of similarity of objects is given in the feature space. It should be noted that the similarity is not the magnitude of the inverse metric. We confine ourselves to constructing a formal definition of clusters in terms of the metric. Let *D* be the matrix

$$D = (d_{i,j})_{k^2}$$

of the pairwise distances between objects of set *B*. We choose a certain threshold δ > 0 and construct a matrix $D_\delta$ whose elements are 1 and 0 in accordance with the following rule:

$$d_{i,j}^{\delta} = \begin{cases} 1, \text{ if } d_{i,j} \leq \delta \\ \\ 0, \text{ otherwise.} \end{cases}$$

Matrix $D_\delta$ is the incidence matrix of some graph $G_\delta$.

Let us consider the problem of covering graph $G_\delta$ with complete edge-disjoint subgraphs. Since the number of cover variants is of order $\mathbf{O}(2^{ck})$, the problem of choosing the best cover variant arises. For this purpose, let us consider the functional

$$J_\delta = \sum_{s=1}^{l} \frac{1}{|K_s|} \sum_{i,j \in K_s} f_s(v_{i,j})$$

where $f_s(v_{i,j})$ is a weight function defined on the set of edges of a complete graph $K_s$, whose number of vertices is $|K_s|$ and $l$ is the number of non-overlapping complete subgraphs of the graph $G(V_\delta, E_\delta)$. We will consider the entire variety $P$ of coverings by complete edge-disjoint subgraphs $\{K_1, K_2,…,K_s\}$. From all covers, we will choose the one for which the functional takes the greatest value. The sets of vertices of complete subgraphs included in the cover of $G(V_\delta, E_\delta)$, which maximizes functional $J_\delta$, will be called the base system of clusters of level $\delta$. As a result of maximizing functional $J_\delta$ for a given argument, we obtain a set of complete subgraphs whose vertex sets form a partition of the set of objects $\{B_1, B_2,…,B_k\}$ into subsets. Such subsets will be called clusters.

Each complete graph includes a set of objects whose pairwise distances are less than or equal to $\delta$. This means that the elements of this set are inside or on the surface of a hypersphere of radius $\delta$. However, it is possible to obtain information on the structure of this set. By structure, we mean the law of the distribution of objects within the boundaries of the mentioned hypersphere. In the meaningful sense of the law of distribution, we mean the degree of uniformity of the distribution of objects inside and on the surface of the hypersphere. However, this information is not enough because it does not reflect the spatial distribution of objects relative to each other. Information of this kind can be represented by structural graphs. Consider a class of structural graphs. A graph will be called structural if it is isomorphic to a geodetic graph and $k$-structural if it is isomorphic to a $k$-geodetic graph [4, 5]. The property of geodeticity reflects the fact that between each pair of vertices there is only one shortest path. Geodetic graphs have remarkable structural properties. They can be expanded or compressed while maintaining their property of geodeticity [7]. Such procedures are associated with the transformation called *homeomorphism*. This transformation preserves the basic structural properties of clusters and classes.

Let us consider several procedures for constructing and analyzing δ-clusters. In the simplest form, each cluster is represented by a complete graph, each of which is obviously geodetic. If the threshold δ is given, then graph *G* describes the structure of the classes in terms of clusters represented by complete graphs. To describe the structure of classes, it is necessary to construct a cover of graph *G* by maximal complete subgraphs. In general, the number of cover options can be very significant. It depends on the choice of the functional of the cover quality, which is usually multi-extremal.

The formulated principle of local compactness assumes that when constructing clusters, it is known the maximum value of δ by which all objects located at a distance of no more than δ from the cluster nucleus will belong to the same cluster. The set of all clusters defined by the principle of δ-compactness determines the generating set of clusters. Classes in automatic classification are the union of clusters from the generating set of clusters.

If the forming clusters have the structure of complete graphs, then the general concept of the cluster remains structurally undefined. As a generalized cluster we mean a geodetic subgraph of a graph *G* of diameter $\leq d$ which is homeomorphic to a complete graph or to a Moore graph. The diameter of the generalized clusters is chosen on the basis of solving problems of maximizing the quality criterion of graph covering by geodetic subgraphs of the two mentioned types.

It should be admitted that an exact cover of a graph *G* by geodetic subgraphs homeomorphic to a complete graph or a Moore graph can be fundamentally impossible for two reasons:

1) graph *G* cannot contain subgraphs isomorphic to the indicated graphs,

2) when covering, there may be edges that are not included in any subgraph homeomorphic to a complete graph or a Moore graph under any cover variant.

Because of this, the concept of a geodetic graph is not enough to study the structure of clusters and classes.

To expand the possibilities for a deeper description of the structures of clusters and classes, it becomes necessary to use *k*-geodetic graphs that reflect the fact that several shortest paths can exist between two points of a cluster or class. Their existence from a content point of view means that they are all inside or on the surface of a certain hypersphere that covers the cluster. *K*-geodetic graphs provide information about the internal structure of a cluster or class that includes a given cluster.

Thus, the structure of generalized clusters is described in the general case by *k*-geodetic graphs homeomorphic to complete graphs or Moore graphs.

It should be noted that a Moore graph possesses the property that there is always a hypersphere of radius δ on the surface of which its vertices are located.

The class structure is described by geodetic graphs that include complete graphs as components that are linked together by edges or common vertices that reflect the spatial location of clusters relative to each other.

Class is a subset of generators of complete subgraphs that represent geodetic subgraphs. The problem of structural classification is reduced to the construction of a cover of graph *G* with vertex-disjoint geodetic subgraphs for which the given functional reaches a maximum. The construction of covering algorithms requires complete information about the structure of geodetic graphs. Article [7] is devoted to describing the structure of all classes of geodetic graphs and enumerating the class of all geodetic graphs homeomorphic to a given geodetic graph. The homeomorphism of geodetic graphs is of special importance since it can be used to describe all variants of proximate structures obtained by means of a unique operation of expansion (compression) of graphs with the preservation of the principle of local compactness.

### 3. The problem of structural correlation analysis based on the theory of geodetic graphs.

The state of many objects of control is described by multidimensional random variables and by both one-dimensional and multidimensional random processes. A set of parameters for the description of control objects can include deterministic variables which, in principle, are regarded as random because they are always measured with errors that are random in nature. The development of mathematical models of objects, the construction of optimal control systems, the forecasting of the dynamics of changes in the states of control objects in time, as a rule, widely use information on the stochastic dependence between all pairs of components of multidimensional random variables. In the case of random processes, the sources of such information are autocorrelation matrices, and for multidimensional random variables, information on stochastic dependences is represented by multidimensional matrices of stochastic dependence.

We restrict ourselves to the case when the correlation coefficient is used as a measure of stochastic dependence. Information about the stochastic dependence between the components of multivariate random variables or the values of a random process at different times is presented in the form of correlation matrices. We can assume that one variant of information processing is reduced to the analysis of the correlation matrix. The analysis is related to the solution of the following classes of applied problems:

a) construction of regression models;

b) selection of significant variables in regression models;

c) identification of groups of closely correlated parameters;

d) description of the structure of the correlation matrix in terms of maximal subsets of independent or quasi-independent parameters and extraction of maximal subgroups of closely correlated parameters;

e) construction of generalized factors in factor analysis.

Let us consider the class of linear regression models of the form

$$y = \sum_{i=1}^{n} a_i x_i + \varepsilon$$

where $y$ is a random variable, vector $(x_1,\ldots, x_n)$ can be either deterministic or random, $\varepsilon$ is a random noise term, and $(a_1,\ldots, a_n)$ is the vector of unknown coefficients. On the basis of the sample matrix $X = (x_{i,j})_{N \times n}$ the estimation of the vector $\hat{\mathbf{a}} = (\hat{a}_1, \ldots, \hat{a}_n)$ is obtained from expression

$$\hat{\mathbf{a}} = (X^T X)^{-1} X^T Y \qquad (1)$$

where $Y^T = (y_1, \ldots, y_n)$ is the transposed vector-column of the sample values of random variable $\varepsilon$. In the calculation of $\hat{\mathbf{a}}$ according to expression (1), a number of computational problems arise that have a profound content meaning. If $\det(X^T X) = 0$, then the estimation of the vector $\hat{\mathbf{a}}$ becomes impossible.

From the content point of view, this means that at least one variable $x_i$ is a linear combination of some subset of the others. Note that there may be several variables, each of which is linearly expressed in terms of the corresponding subsets of other variables. If $\det(X^T X) = 0$, then it is customary to talk about multicollinearity, or modify the computational scheme. The most widespread ones are robust models of estimation in accordance with the following expression:

$$\hat{\mathbf{a}} = (X^T X + \alpha I)^{-1} X^T Y$$

where $\alpha$ is a chosen constant and $I$ is the identity matrix.

$\hat{\mathbf{a}}$ estimations are biased. Bias can be very significant, which leads to a loss of meaningful interpretability of model's coefficients, and as a whole, the model itself also loses concise meaning. Although robust methods of estimating the coefficients of regression models have become widespread, their content value in some cases is not high. Therefore, it is of interest to eliminate the multicollinearity effect by choosing a subset $\{x_i, \ldots, x_{ik}\}$ of set $\{x_1,\ldots,x_n\}$ variables on which the matrix of sample values $X$ no longer gives the multicollinearity effect. The construction of the subset $\{x_i, \ldots, x_{ik}\}$ is based on solving the problem of selecting significant variables in regression models. Variable $x_i$ will be called significant if its removal maximizes the variance of the random variable $\varepsilon$. However, any other variable possessing the same property can turn out to be a linear combination of other significant variables.

Thus, the choice of significant variables in one or another aspect does not guarantee the elimination of the phenomenon of multicollinearity. At the same time, it is obvious that the chosen significant variables should be slightly correlated with the already chosen variables. Such a directed choice of significant variables can be ensured only if there is information about groups of closely correlated variables. A cover of set $\{x_1,\ldots,x_n\}$ by subsets of closely correlated variables is one of the problems of correlation analysis.

The concept of a group of closely correlated variables requires formalization since this will allow us to settle many problems of correlation analysis. Let $\delta > 0$ be a certain threshold and let $R$ be the correlation matrix of the variables $\{x_1,\ldots,x_n\}$. Construct the matrix

$$R_\delta = (r_{i,j}^\delta)_{n \times n}$$

$$r_{i,j}^\delta = \begin{cases} 1, \text{ if } |r_{i,j}| > \delta \\ \\ 0, \text{ otherwise.} \end{cases}$$

Obviously, $R_\delta$ is the adjacency matrix of the vertices of graph $G(V, E_\delta)$, in which the set of vertices is put into a one-to-one correspondence with the set of variables $\{x_1,\ldots,x_n\}$. The graph $G(V, E_\delta)$ can be considered as an ordinary graph. The maximal complete subgraphs of graph $G$, which are geodetic from a content point of view, form a set of variables with a high absolute value of the values of the correlation coefficient. This means that if it exceeds a certain threshold, then such a group of variables has significant redundancy. In the group of variables generated by a complete graph, it is possible to single out one representative type, which reflects the bulk of the information contained throughout the group. We will construct the cover of the graph with complete subgraphs and their representative types. A set of representative types of the constructed cover gives a set of informative variables that are practically orthogonal with respect to each other, and this practically excludes the possibility of the appearance of multicollinearity.

## 4. Conclusions

In Chapter 3 of publication [8] it is analyzed the problem of correlation between structural and spectral metrics in complex networks. In that chapter, it is investigated the Pearson correlation coefficient between extensively studied structural and spectral network metrics. It is pointed out that numerous metrics have been introduced to quantify distinct characteristics of complex networks. Because of the high computational complexity of network metrics, such metrics can be strongly correlated in certain types of complex networks indicating redundancy among them. For this reason, it would be helpful if it was possible to define a small representative set of network metrics that effectively features a given type of complex network. We firmly believe that the same approach to the solution of the problems described in this paper can be used in the solution of the fundamental problems related to strong linear correlations between centrality metrics of complex networks. It is clear that correlations between centrality metrics in complex networks are not well understood yet. One fundamental problem in complex networks is to understand the essential relations between network metrics. The techniques described in this paper could be of great help the purpose of the solution of this fundamental problem.